\documentclass[final]{aipproc}
\layoutstyle{8x11single}

\usepackage{color}

\begin{document}

\title{Introducing TAXI: a Transportable Array for eXtremely large
  area Instrumentation studies}


\author{T.~Karg}{address={DESY, Zeuthen, Germany}}

\author{A.~Haungs}{address={Karlsruhe Institute of Technology,
    Institut f\"ur Kernphysik, Karlsruhe, Germany}}

\author{M.~Kleifges}{address={Karlsruhe Institute of Technology,
    Institut f\"ur Prozessdatenverarbeitung und Elektronik, Karlsruhe,
    Germany}}

\author{R.~Nahnhauer}{address={DESY, Zeuthen, Germany}}

\author{K.-H.~Sulanke}{address={DESY, Zeuthen, Germany}}

\begin{abstract}
  A common challenge in many experiments in high-energy astroparticle
  physics is the need for sparse instrumentation in areas of
  100~km$^2$ and above, often in remote and harsh environments. All
  these arrays have similar requirements for read-out and
  communication, power generation and distribution, and
  synchronization. Within the TAXI project we are developing a
  transportable, modular four-station test-array that allows us to
  study different approaches to solve the aforementioned problems in
  the laboratory and in the field. Well-defined interfaces will
  provide easy interchange of the components to be tested and easy
  transport and setup will allow in-situ testing at different
  sites. Every station consists of three well-understood 1~m$^2$
  scintillation detectors with nanosecond time resolution, which
  provide an air shower trigger. An additional sensor, currently a
  radio antenna for air shower detection in the 100~MHz band, is
  connected for testing and calibration purposes. We introduce the
  TAXI project and report the status and performance of the first TAXI
  station deployed at the Zeuthen site of DESY.
\end{abstract}

\maketitle

\section{Introduction}

The measurement of charged cosmic rays and astrophysical neutrinos at
the highest energies requires extremely large instrumented areas due
to the low flux levels. Ultra-high energy cosmic rays (UHECR) and
neutrinos are correlated due to the fact that the interaction of these
protons with the cosmic microwave background radiation will produce a
guaranteed flux of cosmogenic neutrinos \cite{Berezinsky:1969}. Hence,
the detection of those neutrinos will open a new window to
astrophysics, cosmology, and neutrino physics at high center-of-mass
energies.  UHECR detectors operating today, the surface detectors of
the Pierre Auger Observatory \cite{Allekotte:2008} and of the
Telescope Array experiment \cite{Abu-Zayyad:2012}, already instrument
areas of the order of thousands of $\textrm{km}^2$. Next-generation
detectors with sizes larger than $10\,000~\textrm{km}^2$ are under
study \cite{Blumer:2010}. Recently, a $100~\textrm{km}^2$
low-threshold air shower detector at the South Pole has been proposed
\cite{IceCube:2013} as an atmospheric muon veto for the IceCube
Neutrino Observatory. Two $100~\textrm{km}^2$ experiments aiming at
the detection of cosmogenic neutrinos using radio techniques in ice
are currently under construction: ARA \cite{Allison:2011} and ARIANNA
\cite{Barwick:2007}; large-area hybrid detectors combining optical,
radio, and acoustic detection channels have been proposed
\cite{Vandenbroucke:2006}.

All these detectors have the need for communication between detection
units, low maintenance, decentralized power generation, clock
distribution and trigger generation in common. They are built in harsh
environments, from Antarctic climate to deserts with large daily
temperature variations. Further, for site selection campaigns,
long-term background measurements and signal propagation / detection
studies have to be performed in-situ at different candidate sites.
One project dedicated to study power generation in the field, mainly
in Antarctica, but also at other locations is the ARA autonomous
renewable power station \cite{Besson:2014}. We aim to generalize the
approach of separating the physics detector from the underlying
infrastructure requirements even further in the TAXI project: a
Transportable Array for eXtremely large area Instrumentation studies.

\section{The TAXI Concept}

The goal of the TAXI project is to design and build a modular,
autonomous detection station using well-understood reference air
shower detectors and the possibility to connect any new type of sensor
with waveform readout up to $180~\textrm{MHz}$ sampling rate. Possible
options for the sensor include radio antennas for air shower or
neutrino detection, muon detectors of any type or hybrid
(radio/muon/electron) detectors for air shower detection. Also
photomultiplier tubes (PMTs) for non-imaging Cherenkov telescopes, or,
at lower sampling rates, acoustic detectors can be
connected. Transient electromagnetic signals in the GHz band can be
recorded using signal envelope techniques.  Well-defined power and
communications interfaces will allow us to test different approaches
to power generation and data transfer in the field. With a single
station new sensors can be tested and calibrated with a known air
shower trigger with excellent timing and basic directional
sensitivity. A small array of four TAXI stations enables the
development and long-term in-situ testing of new communication and
power generation systems under realistic trigger conditions. This
leads to the following requirements for TAXI:

\begin{description}
\item[High modularity] allows easy interchange of components,
  especially the power supply and communication modules. The
  interfaces will be defined up to the connector types and pinouts
  including the underlying communication protocol.
\item[Easy transport and setup] allows site studies for future
  projects using custom sensors with full waveform readout. This
  includes long-term monitoring of backgrounds and in-situ signal
  propagation studies (signal speed, attenuation, refraction, and
  others).
\item[Operation at isolated sites] requires low power consumption and
  a self-sustaining power supply like photovoltaic, wind, or similar
  systems. Further, the system shall operate in the full range of
  Antarctic to hot climate zones.
\end{description}

\begin{figure}
  \includegraphics[height=0.3\textwidth]{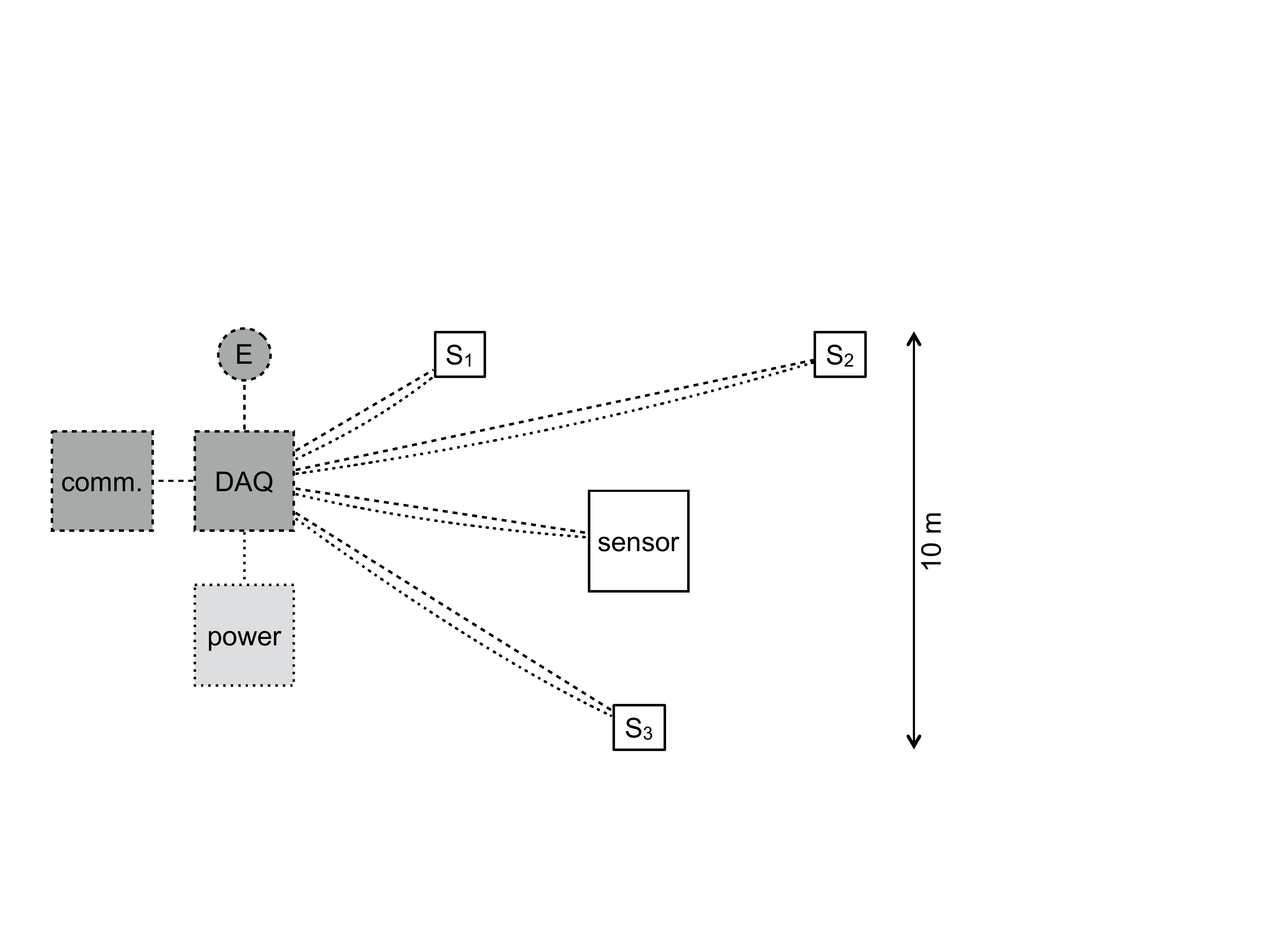}
  \caption{Schematic drawing of a single TAXI station: it comprises three
    scintillator reference detectors $\textrm{S}_i$ deployed at
    typically $10~\textrm{m}$ mutual distance, a test sensor,
    power (dotted) and communication (dashed) links, and environmental
    monitoring $E$.}
  \label{fig:station_schematic}
\end{figure}

\noindent Figure~\ref{fig:station_schematic} shows schematic drawing
of a single TAXI station.

\section{Design and Performance of the First TAXI Station}

A first TAXI prototype station has been constructed and has been
deployed on the Zeuthen site of DESY in June 2013. As air shower
reference detectors segmented $1~\textrm{m}^2$ plastic scintillator
plates with nanosecond time precision \cite{Bahr:1998} are used. In
each plate four $50 \times 50~\textrm{cm}^2$ segments are read out
separately. Every segment is read out with two PMTs in coincidence
mode for noise suppression; two Hamamatsu R5900-03-M4 four-channel
PMTs are installed for this purpose. The scintillator plate, PMTs, and
HV generation are housed in a weather-proof aluminum housing that is
supplied with a voltage of $\pm 12~\textrm{V}$. The eight PMT signals
are transmitted via differential signalling over shielded twisted pair
cables to the station DAQ.

In the first prototype station, the PMT signals are processed by a
VME-based DAQ system, which is read-out via USB by a Raspberry
Pi\footnote{\url{http://www.raspberrypi.org/}} single board
computer. The signals of all 12 segments (three plates, four segments
each) are split and are i) discriminated and recorded by a
time-to-digital converter (TDC) with $1~\textrm{ns}$ time resolution
and ii) integrated by a charge sensitive ADC for calibration and
monitoring purposes.

The discriminated signals are routed to a custom-made trigger board. A
FPGA on the trigger board allows programming arbitrary trigger
conditions between the 12 segments. Currently we require at least one
segment in each of the three scintillator plates to record a signal
above threshold in a $400~\textrm{ns}$ time window.

For the waveform readout of the test sensor the digitizer board
developed for the Auger Engineering Radio Array (AERA)
\cite{Berg:2013} is used (see e.g.~\cite{Schmidt:2012} for a detailed
description of the board). For TAXI we currently employ two of the
four $180~\textrm{MHz}$ ADC channels with a buffer depth of 7
seconds. On an external trigger signal from the DAQ of the
scintillation detector, $11.4~\mu\textrm{s}$ (2048 samples) of
waveform data around the trigger time are stored together with the
scintillation detector event and the data are timestamped using the
GPS module integrated on the AERA board. As a test sensor we currently
use a Short Aperiodic Loaded Loop Antenna (SALLA) for radio air shower
detection from the Tunka-Rex experiment \cite{Hiller:2013}, originally
developed for the LOPES experiment \cite{Kromer:2009}. The E-W and N-S
polarization are read out separately by the two ADC channels on the
AERA digitizer board.

Environmental monitoring is performed with a commercial weather
station. In addition, temperature and humidity sensors are installed
in the central DAQ housing. Further, the temperature of the Raspberry
Pi single board computer and of the AERA digitizer board are monitored
continuously.

The first TAXI prototype station is operating without problems since
deployment; more than one year of data have been collected now. The
thresholds for the scintillation detector are set to a trigger rate of
$40~\textrm{Hz}$ for each segment after the coincidence of the two
PMTs reading out the segment. Requiring at least one triggered segment
in each of the three plates this results in a global trigger rate of
approx.~$1~\textrm{min}^{-1}$.

\begin{figure}
  \includegraphics[width=0.45\textwidth]{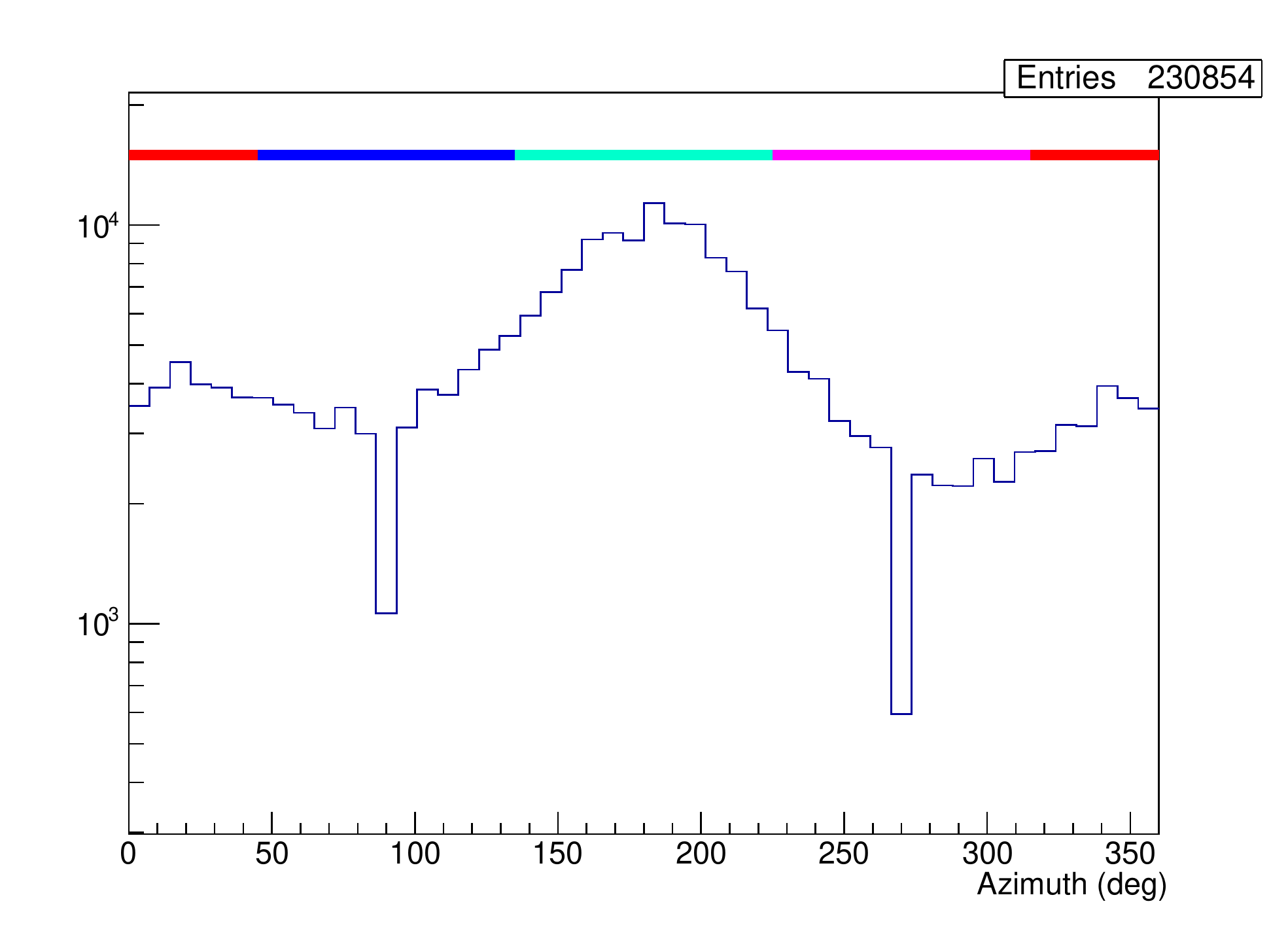}
  \hfill \includegraphics[width=0.45\textwidth]{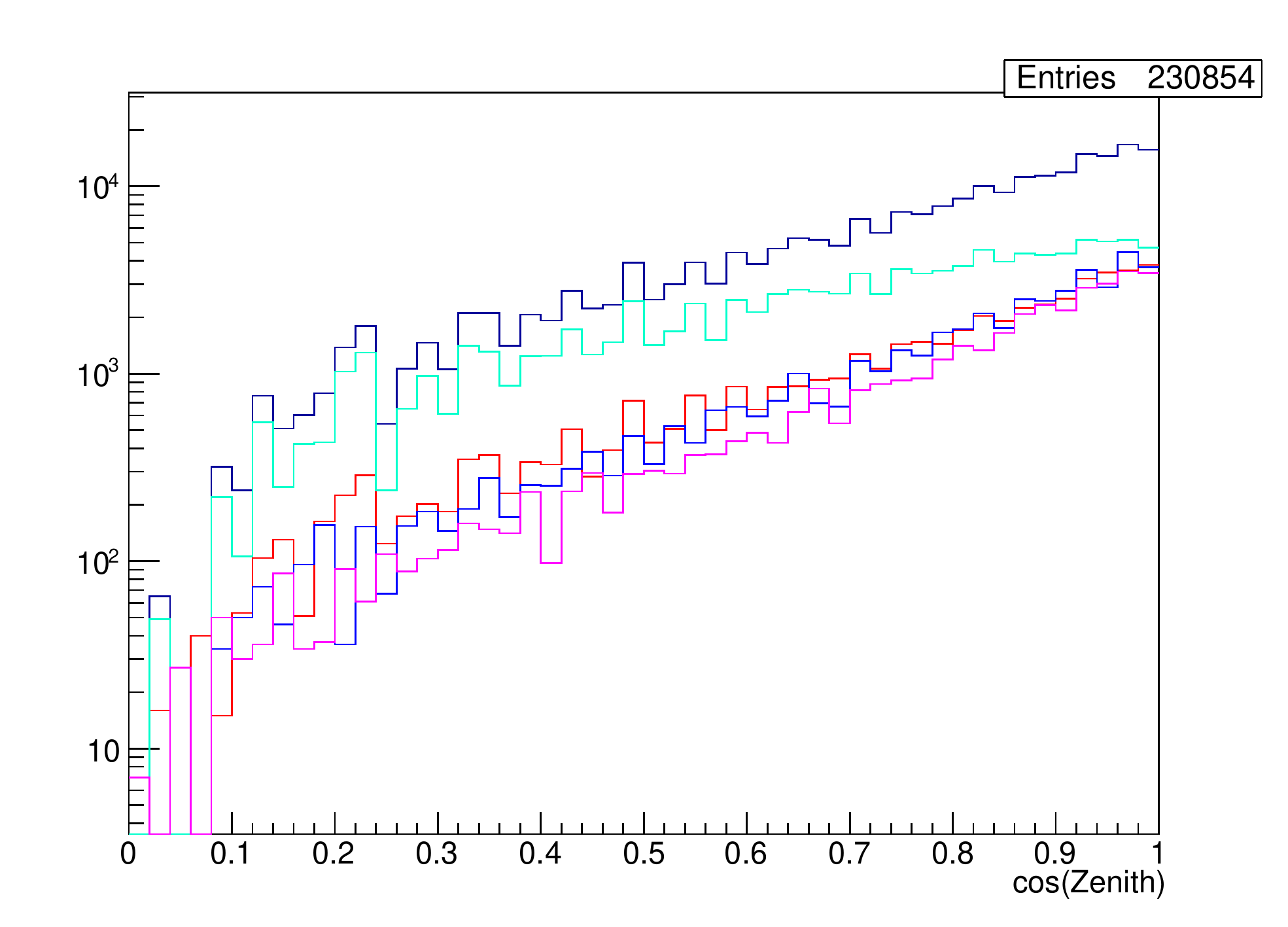}
  \caption{Air shower directions reconstructed from the arrival time
    differences in the scintillation reference detectors. The colored
    zenith-histograms correspond to the four different azimuthal
    ranges indicated by the horizontal band in the azimuth
    distribution.}
  \label{fig:direction_reconstruction}
\end{figure}

The measurement of the signal arrival time differences in the three
scintillator plates with the TDC allows the reconstruction of the
direction of the air shower assuming a plane wave front. The
reconstructed arrival directions for all showers where the
corresponding system of linear equations had a solution, are shown in
Fig.~\ref{fig:direction_reconstruction}. The deficits visible at
azimuthal angles of $90^\circ$ and $270^\circ$ are an artefact of the
reconstruction algorithm used. The colored zenith-histograms
correspond to four different azimuthal ranges as indicated by the band
in the azimuth histogram. It can be seen that for vertical showers the
rate in all azimuthal ranges is equal while in three azimuthal sectors
the shower rate is suppressed for inclined showers. This structure can
be explained by the shadowing through buildings surrounding the TAXI
prototype station on the roof of the mechanical workshop at the
Zeuthen site of DESY.

\section{Next Steps}

\begin{figure}
  \centering
  \includegraphics[width=0.82\textwidth]{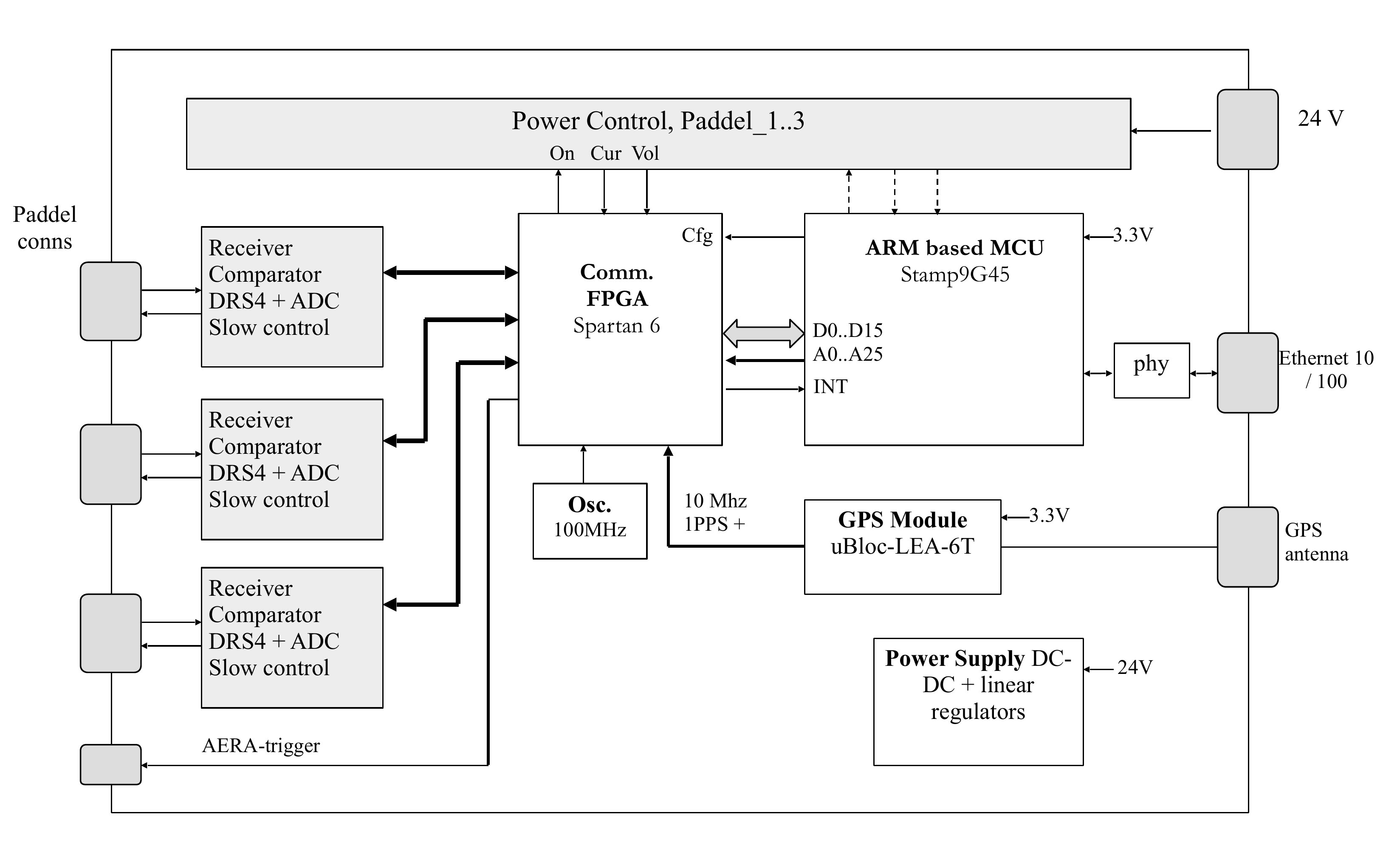}
  \caption{Block diagram of the TAXI readout electronics currently
    being developed.}
  \label{fig:readout}
\end{figure}

To meet the low-power requirement the VME-based DAQ system for the
scintillation detector will be replaced by a new single-board DAQ
system that will incorporate the TDC functionality, FPGA based
triggering, monitoring, and slow control. In addition, it will
optionally offer the possibility to record PMT waveforms from the
scintillators for monitoring and calibration purposes via a DRS4
switched capacitor array \cite{Ritt:2008}. Figure~\ref{fig:readout}
shows a block diagram of the TAXI readout board currently under
development. The TDC functionality will be realized in the FPGA with
an expected precision of $1~\textrm{ns}$. With the DRS4 and
corresponding ADCs powered off, the TAXI readout board will require
less than $10~\textrm{W}$ of power which can be supplied in the field
by solar panels or wind turbines.

We expect to construct and install four TAXI stations using the new
readout board by the end of 2014. This first mini-array will use solar
power and commercial WiFi for communication.

\begin{theacknowledgments}
  This work is supported by the ``Helmholtz Alliance for Astroparticle
  Physics HAP'' funded by the Initiative and Networking Fund of the
  Helmholtz Association.
\end{theacknowledgments}

\bibliographystyle{aipproc}
\bibliography{karg}

\end{document}